\documentstyle[sprocl,epsfig]{article}

\def\be{\begin{equation}}
\def\ee{\end{equation}}
\def\bea{\begin{eqnarray}}
\def\eea{\end{eqnarray}}

\begin{document}

\title{Heavy Quarks at Threshold: Recent Developments\\}

\author{Matthias Steinhauser}

\address{II. Institut f\"ur Theoretische Physik, Universit\"at Hamburg,\\ 
  22761 Hamburg, Germany}


\maketitle\abstracts{
Recent developments in the treatment of a system of heavy quarks
close to their production threshold are discussed. Particular emphasis
is put on the next-to-next-to-next-to-leading order (N$^3$LO) corrections
to the energy levels and the wave function and their
effect on the cross section $\sigma(e^+e^-\to t\bar{t})$. Furthermore,
the resummation of logarithmic corrections within the framework of
potential non-relativistic QCD (pNRQCD) are considered.
}


\vspace*{-2em}
\section{Introduction}

One of the most important tasks of a future linear collider is the
precise measurement of the cross section for the production of top quark pairs
close to their production threshold. Next to the mass and the width of the top
quark also the strong coupling and --- in case the Higgs boson is not too
heavy --- also the top quark Yukawa coupling can be determined with a quite
high accuracy. In order to match the expected experimental
precision~\cite{Martinez:2002st}  
it is important to compute higher-order quantum corrections to this process.

Due to the large mass and width top quarks do not form bound states. Instead,
the resonance curve in the threshold region shows a local maximum below the
nominal threshold which is essentially the remnant of a would-be $1S$ bound
state.

The NNLO corrections to the cross section $e^+ e^- \to
t\bar{t}$ in the threshold region has been completed a few years ago by four 
independent groups. The comparison of the individual results is discussed
in Ref.~\cite{Hoa_etal}. The essential outcome of the combined effort is that
no stability of the position of the peak is found in case the pole quark mass
is used for the parametrization of the cross section. Thus a precise
determination of the quark mass would not be possible.
However, this changes if one switches to a short-distance mass which leads to
a stabilization of the peak position and thus to a reliable
extraction of the corresponding mass parameter.
The normalization in the peak position, which is essenially determined
by the wave function at the origin, still remains unstable after including NLO
and NNLO corrections.

Shortly after the completion of the second order calculation there have been
attempts to go beyond the fixed-order approximation and resum the logarithms
in the velocity of the top quark. In Ref.~\cite{HMST} an
accuracy of 2-3\% was claimed for the cross section of $t\bar t$ threshold
production. However, subsequent calculations of further  
next-to-next-to-leading logarithmic (NNLL) terms \cite{Hoa}, which had not
been taken into account in Ref.~\cite{HMST}, casted serious doubts on this
optimistic estimate. Thus, the full calculation of the NNLL corrections, which
still remains elusive, is unavoidable to draw definite conclusions. 

In these proceedings we want to report about phenomenological applications 
which are based on higher order results obtained within the framework of
pNRQCD~\cite{PinSot1}.
In particular, it was possible to evaluate the fixed-order N$^3$LO
corrections to the ground state energy~\cite{KPSS1,PenSte}. 
After the inclusion of the third-order corrections an impressive
stabilization in the position of the peak is observed --- even for
on-shell quark masses. Thus this result can be used
to determine the bottom quark mass from the $\Upsilon(1S)$ system
and the top quark mass from the position of the peak
in the cross section~\cite{PenSte}. The latter will be
discussed in Section~\ref{sec::topmass}.
Futhermore, the quadratic~\cite{KniPen2} and linear~\cite{KPSS2,Hoa}
logarithms
of the third-order corrections to the wave function have been determined.
The inclusion of these terms seems to provide more stability
in the normalization of the peak.
However, for a definite conclusion, also the
constant term at N$^3$LO is needed.

In the meantime first results containing the summation of higher-order
logarithms within pNRQCD became available for the spin-dependent part
of the Hamiltonian. In particular the NLL corrections to the hyperfine
splitting (HFS)\footnote{Note that this corresponds to the N$^3$LL corrections
to the spin-dependent part of the energy levels.} 
have been evaluated in~\cite{KPPSS}. In Section~\ref{sec::etab} we will
discuss how this result can be used to predict the mass of the
$\eta_b$ meson.
In Ref.~\cite{PPSS} the results of~\cite{KPPSS} have been generalized to the
non-equal mass case and applied to the HFS in the $B_c$ system.
For completness we want to mention that 
recently the NNLL corrections to the 
ratio of the spin-1 and spin-0 mediated production and 
annihilation processes of heavy quarks have been computed~\cite{PPSS2}.


\vspace*{-.5em}
\section{\label{sec::topmass}Top quark mass determination}

For the top quark system the relation between the 
resonance energy and the quark mass is given by
\begin{eqnarray}
  E_{\rm res}&=&2m_t+E_1^{\rm p.t.}+\delta^{\Gamma_t}E_{\rm res}\,,
  \label{ttcorr}
\end{eqnarray}
where $E_1^{\rm p.t.}$ is the perturbative contribution to the ground state
energy and $\delta^{\Gamma_t} E_{\rm res}=100\pm 10~{\rm MeV}$
is the effect of higher-order resonances and the finite width of the top
quark. Non-perturbative effects are negligible.

\begin{figure}[t]
  \begin{center}
    \begin{tabular}{c}
      \leavevmode
      \epsfxsize=7.0cm
      \epsffile[40 230 550 580]{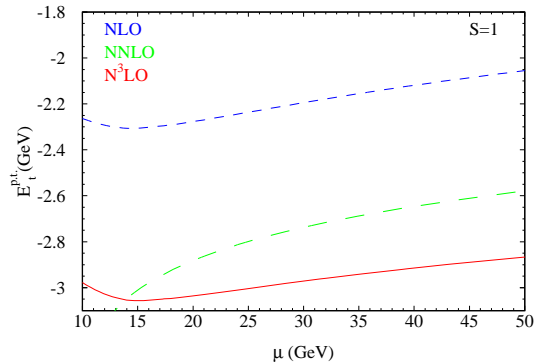}
    \end{tabular}
  \end{center}
  \vspace*{-1em}
  \caption{\label{fig:top1}
  Ground state energy $E^{\rm p.t.}_1$ of the 
  $t\bar{t}$ bound state in the zero-width approximation 
  as a function of the renormalization scale $\mu$
  for $S=1$.
  The short-dashed (blue), long-dashed (green) and solid (red) line 
  corresponds to the NLO, NNLO and N$^3$LO approximations.
          }
\end{figure}

To evaluate  the perturbative contribution we use
the result of Ref.~\cite{PenSte} for $E_1^{\rm p.t.}$
up to N$^3$LO. In Fig.~\ref{fig:top1} $E_1^{\rm p.t.}$ is plotted
in NLO, NNLO and N$^3$LO approximation as a function of the 
renormalization scale of the strong coupling constant
for $S=1$.
One can see, that the N$^3$LO result shows a much weaker 
dependence on $\mu$ than the NNLO one. 
Moreover at the scale $\mu\approx 15$~GeV,
which is  close to the physically motivated soft
scale $\mu_s\approx 30$~GeV,
the N$^3$LO correction vanishes and furthermore 
becomes independent of $\mu$, {\it i.e.} the N$^3$LO curve shows a local
minimum. This suggests
the convergence of the  series for 
the ground state energy in the pole mass scheme.
Similar results for the $\overline{\rm MS}$ quark mass can be found in
Ref.~\cite{Kiyo:2002rr}. 

In Ref.~\cite{PenSte} a universal 
relation has been obtained between the resonance energy in $t\bar t$ threshold
production in $e^+e^-$ annihilation or $\gamma\gamma$ collisions
and the top quark pole mass  
\begin{equation}
  E_{\rm res}= (1.9833 
  + 0.007\,\frac{m_t-174.3~\mbox{GeV}}{174.3~\mbox{GeV}}
  \pm 0.0009)\times m_t\,.
  \label{ttcorrnum1}
\end{equation}
The central value is computed for $m_t=174.3$~GeV.
For the estimate of the theoretical uncertainty in 
Eq.~(\ref{ttcorrnum1})
we  assume a $\pm 100\%$ error in the  Pad\'e approximation  
of the still unknown three-loop static potential
coefficient and vary the normalization scale
in the stability interval which is roughly given by
$0.4\mu_s<\mu<\mu_s$. Furthermore we use $\alpha_s(M_Z)=0.1185\pm0.002$ and
take into account the $\pm10~{\rm MeV}$ uncertainty in 
$\delta^{\Gamma_t} E_{\rm res}$.
Due to the very nice behaviour of the perturbative expansion
for the ground state energy we do not expect large higher 
order corrections to our result. 
We want to mention that $m_t$ extracted from Eq.~(\ref{ttcorrnum1})
leads to a fixed-order quark mass which, by definition, does not suffer from
renormalon ambiguities. Furthermore, it can be converted with high accuracy
to the $\overline{\rm MS}$ quark mass.


\vspace*{-.5em}
\section{\label{sec::etab}Prediction of $M(\eta_b)$}

A compact analytical expression for the HFS to NLL,
$E_{\rm hfs}^{\rm NLL}$, is given in Eq.~(1) of Ref.~\cite{KPPSS}.
$E_{\rm hfs}^{\rm NLL}$ is a
function of $y=\alpha_s(\mu)/\alpha_s(m_b)$.

\begin{figure}
  \begin{tabular}{cc}
      \leavevmode
      \epsfxsize=5cm
      \epsffile[30 40 640 440]{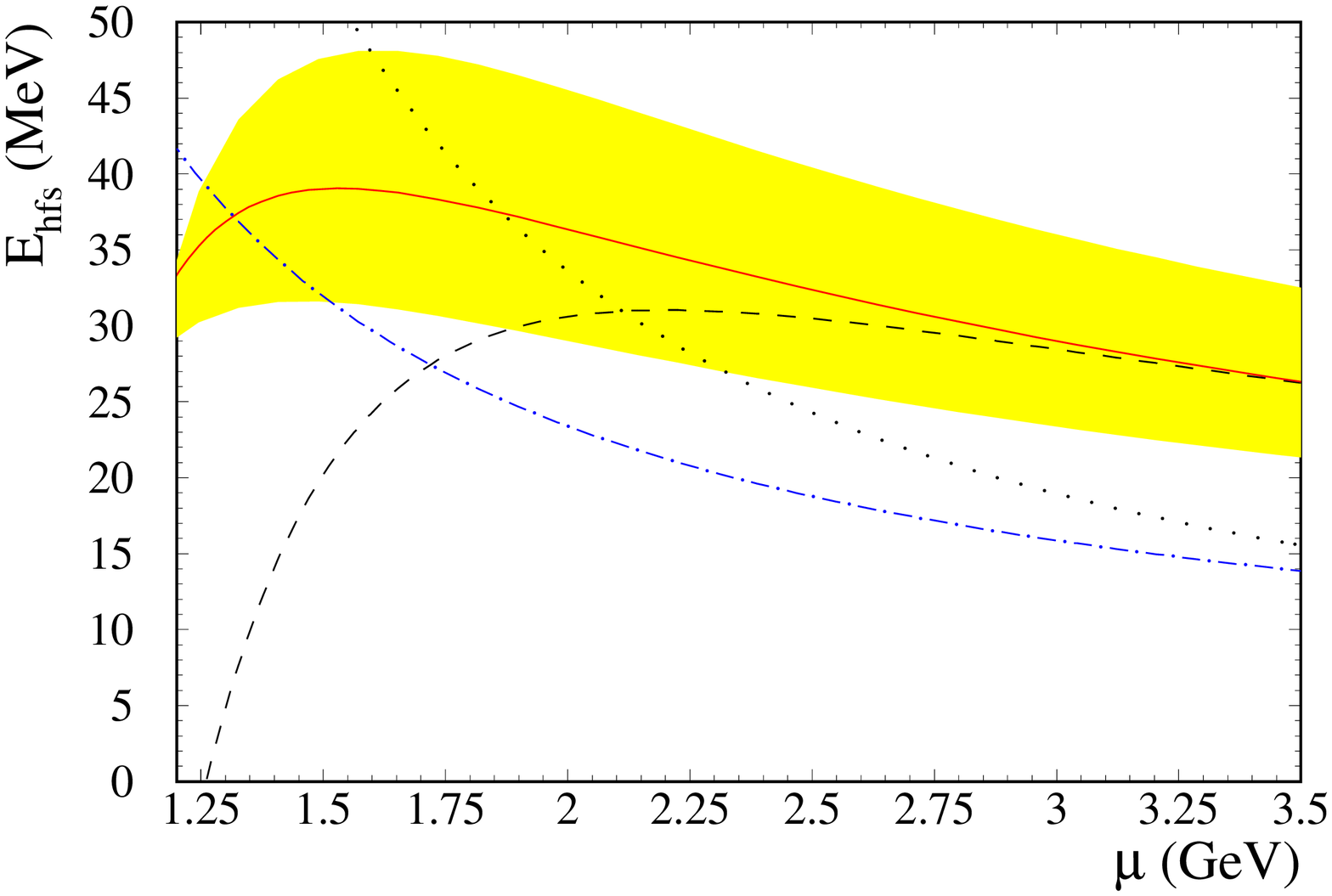}
&
      \leavevmode
      \epsfxsize=5cm
      \epsffile[30 40 640 440]{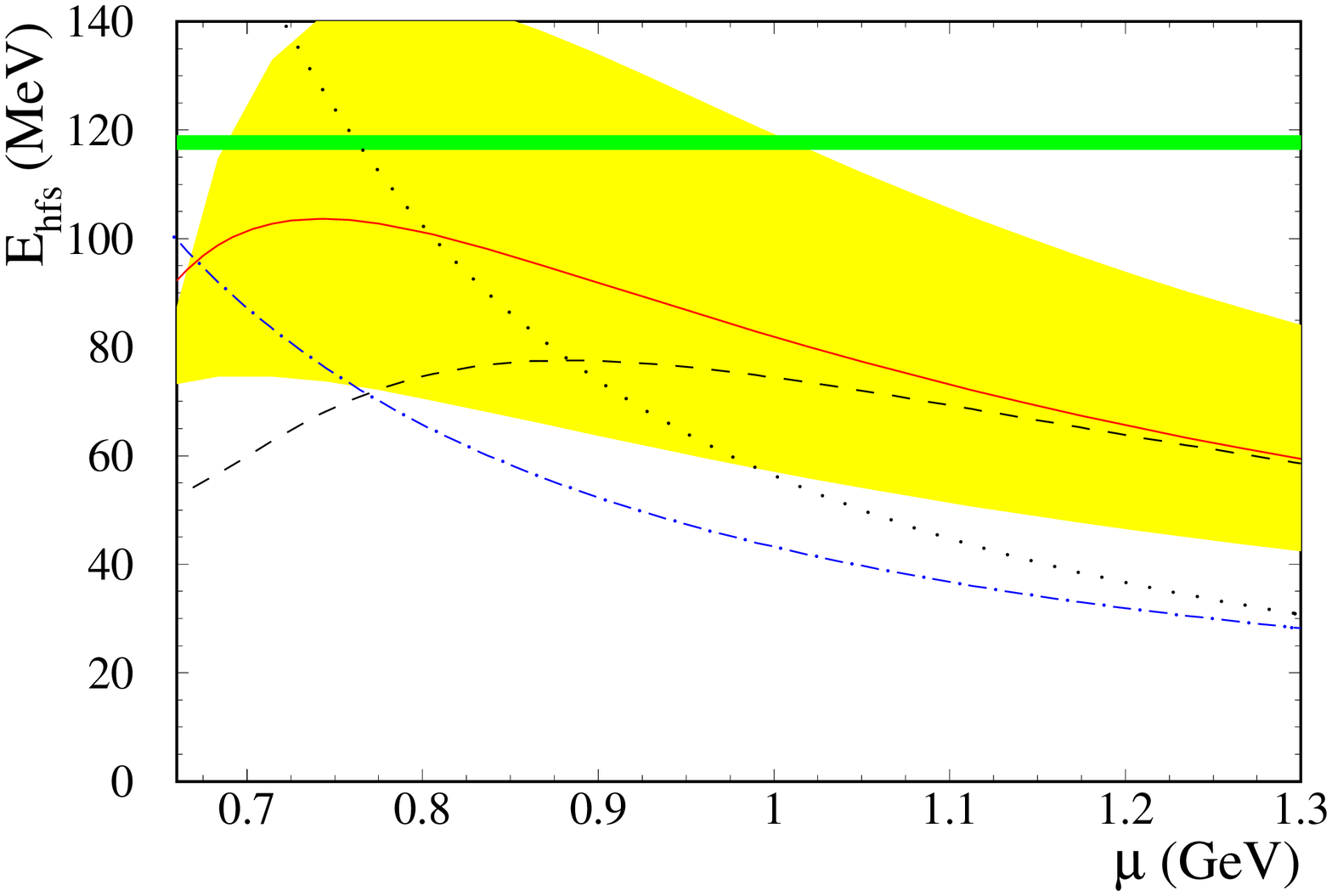}
\\
(a) & (b)
\end{tabular}
  \vspace*{-1em}
\caption{\label{fig1} HFS of 1S bottomonium (a) and charmonium (b)
as a function of the
renormalization scale $\mu$ in the LO (dotted line), NLO (dashed line), LL
(dot-dashed line), and NLL (solid line) approximations. For the NLL
result, the (yellow) band reflects the errors due to $\alpha_s(M_Z)=0.118\pm
0.003$. The horizontal (green) band in (b) gives 
the experimental value $117.7\pm 1.3$~MeV.}
\end{figure}

In Fig.~\ref{fig1}(a), the HFS for the bottomonium
ground state is plotted as a function of $\mu$ in the LO, NLO, LL, and NLL
approximations. As can be seen, the LL curve shows a weaker scale dependence
compared to the LO one. The scale dependence of the NLO and NLL expressions is
further reduced, and, moreover, the NLL approximation remains stable up to
smaller scales than the fixed-order calculation. At the scale $\mu'\approx
1.3$~GeV, which is close to the inverse Bohr radius, the NLL correction
vanishes.  Furthermore, at $\mu''\approx 1.5$~GeV, the result becomes
independent of $\mu$; {\it i.e.}, the NLL curve shows a local maximum.  This
suggests a nice convergence of the logarithmic expansion despite the presence
of the ultrasoft contribution with $\alpha_s$ normalized at the rather low
scale $\bar\mu^2/m_b\sim 0.8$~GeV.  By taking the difference of the NLL and LL
results at the local maxima as a conservative estimate of the error due to
uncalculated higher-order contributions, we get $E_{\rm hfs}=39\pm 8$~MeV. A
similar error estimate is obtained by the variation of the 
normalization scale in the physically motivated soft region $1-3$ GeV.

So far only the perturbative part of the HFS has been discussed.
In Ref.~\cite{KPPSS} it has been argued that the non-perturbative
contributions are small. On one hand this is supported by the good
agreement of our result with recent lattice analysis (see discussion
in Ref.~\cite{KPPSS}). On the other hand, in the charm system 
very good agreement of the HFS with the experimental data is obtained
as can be seen in Fig.~\ref{fig1}(b) where the 
experimental value $117.7\pm 1.3$~MeV is very close to the theoretical value
at the local maximum of the NLL curve which reads
$E_{\rm hfs}=104$~MeV.  
We should emphasize the crucial role of the resummation to
bring the perturbative prediction closer to the experimental
figure.

Our final prediction of the mass of
the as-yet undiscovered $\eta_b$ meson reads
\begin{equation}
M(\eta_b)=9421\pm 11\,{(\rm th)} \,{}^{+9}_{-8}\, 
(\delta\alpha_s)~{\rm MeV}\,,
\end{equation}
where the errors due to the high-order perturbative corrections and the
nonperturbative effects are added up in quadrature in ``th'', whereas
``$\delta\alpha_s$'' stands for the uncertainty in
$\alpha_s(M_Z)=0.118\pm0.003$.  If the experimental error in future
measurements of $M(\eta_b)$ will not exceed a few MeV, the bottomonium HFS
will become a competitive source of $\alpha_s(M_Z)$ with an estimated accuracy
of $\pm 0.003$, as can be seen from Fig.~\ref{fig1}. 


\vspace*{-.5em}
\section*{Acknowledgments}

I would like to thank B. Kniehl, A. Penin, A. Pineda and V. Smirnov for a
very pleasant and fruitful collaboration. This work was supported by
HGF Grant No. VH-NG-008.


\end{document}